\documentclass[prc,aps,floatfix,twocolumn,groupedaddress,nofootinbib,showpacs,preprintnumbers,
amsmath,amssymb,amsfonts,widetable] {revtex4-1}
\usepackage{bm}
\usepackage{soul}
\usepackage{array}
\usepackage{lipsum}
\usepackage{relsize}
\usepackage{mathrsfs}
\usepackage{amssymb}
\usepackage{amsmath}
\usepackage{graphicx}
\usepackage[usenames]{color}
\usepackage{pslatex}
\usepackage{booktabs}

\begin{document}
\title{Applications of reduced basis methods to the nuclear single particle spectrum}
\author{Amy L. Anderson}\email{aanderson6@fsu.edu}
\author{Graham L. O'Donnell}\email{glo15@fsu.edu}
\author{J. Piekarewicz}\email{jpiekarewicz@fsu.edu}
\affiliation{Department of Physics, Florida State University,
               Tallahassee, FL 32306, USA}
\date{\today}
\begin{abstract}
 Reduced basis methods provide a powerful framework for building efficient
 and accurate emulators. Although widely applied in many fields to simplify
 complex models, reduced basis methods have only been recently introduced 
 into nuclear physics. In this letter we build an emulator to study the single-particle 
 structure of atomic nuclei. By scaling a suitable mean-field Hamiltonian, a 
 ``universal" reduced basis is constructed capable of accurately and efficiently 
 reproduce the entire single-particle spectrum of a variety of nuclei. Indeed, 
 the reduced basis model reproduces both ground- and excited-state energies 
 as well as the associated wave-functions with remarkable accuracy. Our results 
 bode well for more demanding applications that use Bayesian optimization to 
 calibrate nuclear energy density functionals.
\end{abstract}
\smallskip

\maketitle

Eigenvector continuation (EC) is a novel method for calculating eigenstates of a Hamiltonian matrix defined 
by one or more variable parameters\,\cite{Frame:2017fah}. The core assumption behind the success of EC 
is that eigenstates vary smoothly over the manifold defined by the parameters, so that the eigenstates for a 
given parameter are likely to be well approximated by a linear combination of the eigenstates obtained with 
another set of parameters. Where a direct calculation of eigenstates---such as direct matrix diagonalization in 
large vector spaces---would be computationally demanding, EC transforms the problem into a simple 
diagonalization in a low-dimensional space. The nuclear physics community has benefited greatly from these 
new insights and has developed a set of accurate and efficient emulators using eigenvector 
continuation\,\cite{Konig:2019adq,Furnstahl:2020abp,Drischler:2021qoy}.

Recently, EC has been identified to belong to a general class of techniques that fall under the general rubric of 
``reduced basis methods'' (RBMs)\,\cite{Quarteroni:2015,Heasthaven:2016}. Although new to nuclear 
physics\,\cite{Bonilla:2022rph,Melendez:2022kid}, reduced basis methods---as part of the general framework 
of reduced order models\,\cite{Benner:2020}---is a relatively mature field that offers efficient and accurate solutions 
to numerically challenging problems over a wide scientific landscape\,\cite{Benner:2020}. It is the goal of this letter 
to continue the application of RBMs to nuclear science, particularly in the context of the independent particle model.

The independent particle model is a fundamental pillar of nuclear structure. As argued by Bohr and Mottelson: ``the 
relatively long mean free path of the nucleons implies that the interactions primarily contribute a smoothly varying 
average potential in which the particles move independently''\,\cite{BohrI:1998}. Self-consistent mean-field models
that are at the core of nuclear energy density functionals exploit this paradigm to find the optimal set of single 
particle orbitals. Even more sophisticated models, such as ab initio no core shell model and coupled cluster
theory often start from a simple single-particle basis that preserves translational symmetry\,\cite{Barrett:2013nh} or
from a reference state that consists of a Slater determinant of single-particle orbitals that gets refined by the inclusion 
of many-body correlations\,\cite{Hagen:2013nca}.

Despite the remarkable advances in algorithmic development, computer power, and physical insights, diagonalizing 
Hamiltonian matrices in model spaces containing millions of basis states is often required. As such, reduced basis 
methods can provide a framework to drastically increase computational speed while retaining accuracy.  RBMs have only 
been recently introduced into nuclear physics, so many questions remain on its applicability to the many difficult 
problems permeating the field. In this work we demonstrate how RBMs can generate a ``universal'' set of basis states 
that may be used to compute the entire single-particle spectrum---both ground and excited states---of a variety of nuclei 
across the nuclear chart\,\cite{ODonnell:2022}. 

To start, we introduce a dimensionless Schr\"odinger equation in the presence of a spherically symmetric 
mean-field Hamiltonian $\hat{H}|{\cal U}_{\,n\kappa}\rangle \!=\! \epsilon|{\cal U}_{\,n\kappa}\rangle$, which
in configuration space turns into the following second-order differential equation:
\begin{equation}
  \left(\!-\frac{d^{2}}{dx^{2}} + V(x) \!+\! 
  \frac{\kappa(\kappa\!+\!1)}{x^{2}}\right)\!{\cal U}_{\,n\kappa}(x) \!=\!
 \epsilon{\cal U}_{\,n\kappa}(x).
 \label{Hamiltonian}
\end{equation}
Here $\epsilon$ is the dimensionless energy to be defined later and $\kappa$ is a shorthand notation for both the 
total angular momentum $j\!=\!|\kappa|\!-\!1/2$ and the orbital angular momentum $l$
\begin{equation}
 l = \begin{cases}
       \kappa             & \text{if } \kappa > 0 \\
      -(1\!+\!\kappa)   & \text{if } \kappa <0.
  \end{cases}
 \label{kappa} 
\end{equation}
Note that $\kappa(\kappa\!+\!1)\!=\!l(l\!+\!1)$. The spherically symmetric mean-field potential $V(x)$ includes central, 
Coulomb, and spin-orbit contributions that are parametrized in terms of a Woods-Saxon potential, a Coulomb potential 
derived from an assumed Gaussian charge distribution, and the derivative of a Woods-Saxon potential, respectively. 
That is,
\begin{equation}
  V(x)=-\lambda_{{}_{0}}f_{{}_{0}}(x)+\lambda_{c}f_{c}(x)+(1+\kappa)\lambda_{\rm so}f_{\rm so}(x),
 \label{Potential1} 
\end{equation}
where
\begin{subequations}
\begin{align} 
 & f_{{}_{0}}(x) = \bigg[1+\exp\Big(\beta(x-1)\Big)\bigg]^{-1}, \\
 & f_{c}(x) = \bigg[\frac{{\rm erf}(x)}{x}\bigg], \\
 & f_{\rm so}(x) = \bigg[x\cosh^{2}\!\Big(\beta(x\!-\!1)/2\Big)\bigg]^{-1}.
\end{align}
\label{Potential2}
\end{subequations}
The only model parameter that appears in these expressions is $\beta\!=\!c/a$, which is defined as the ratio of the
half density radius $c$ to the diffuseness parameter $a$ of the Wood-Saxon potential. With the exception of $\beta$, 
the mean-field potential depends linearly on the three strength parameters, $\lambda_{{}_{0}}$, $\lambda_{c}$, and 
$\lambda_{\rm so}$. The linear dependence of the Hamiltonian on the model parameters is an important condition 
for the efficient performance of reduced basis emulators\,\cite{Heasthaven:2016}.

\begin{figure}[ht]
 \centering
 \includegraphics[width=0.45\textwidth]{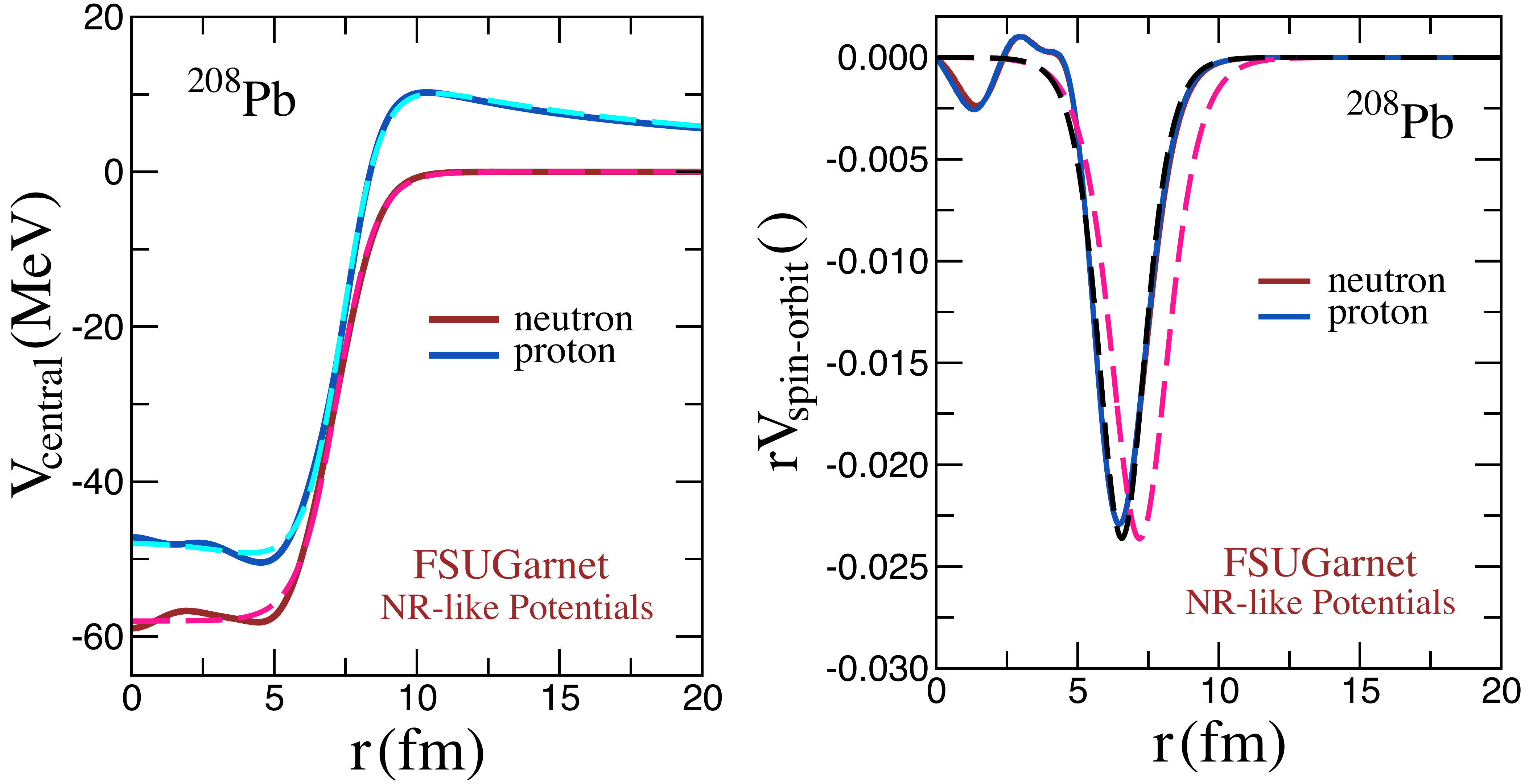}
 \caption{Depicted by the solid lines are neutron and proton effective ``Schr\"odinger-like'' central 
 	       and spin-orbit potentials for ${}^{208}$Pb derived from the covariant energy density 
	       functional FSUGarnet\,\cite{Chen:2014mza}. The dashed lines represent the corresponding
	       Woods-Saxon fits. The slightly offset dashed line represents a fit to the spin-orbit potential 
	       that uses the same half-density radius and diffuseness as the central potential.}
\label{Fig1}
\end{figure}

To determine the various model parameters across the nuclear chart we rely on the predictions of 
FSUGarnet\,\cite{Chen:2014mza}, a realistic covariant energy density functional calibrated to the properties of 
finite nuclei and neutron stars. Effective non-relativistic ``Schr\"odinger-like'' central and spin-orbit potentials naturally 
emerge from such a description\,\cite{Amado:1983wio} and are plotted in Fig.\ref{Fig1} for 
the case of ${}^{208}$Pb.
Without any further adjustments, a Woods-Saxon form and its associated derivative provide a highly accurate 
representation of the non-relativistic potentials. The proton fit to the central potential also includes a Coulomb 
contribution as in Eqs.(\ref{Potential1}-\ref{Potential2}). Note that for simplicity, we have assumed that both the 
half-density radius $c$ and the diffuseness $a$ of the spin-orbit potential are identical to the corresponding 
parameters of the central potential. Finally, we note that the effective Schr\"odinger-like central and spin-orbit 
potentials depend on energy\,\cite{Amado:1983wio}, so we have fixed its value to an average binding energy 
of 40\,MeV. Optimal values for the Woods-Saxon parameters for a variety of spherical nuclei are listed in 
Table\,\ref{Table1}. In this first publication we highlight the power and flexibility of the approach by focusing on 
the neutron single-particle spectrum.

\begin{table*}[ht]
\begin{tabular}{|c|c|c|c|c||c|c|c|}
 \hline\rule{0pt}{2.5ex}  
 Nucleus & $V_{{}_{0}}$ (MeV) & $V_{\rm so}$ & $c$ (fm) & $a$ (fm) & 
 $\lambda_{{}_{\,0}}$ & $\lambda_{\,\rm so}$ & $\beta$ \\ 
\hline\rule{0pt}{2.5ex}  
 $^{16}\mathrm{O}$    & 62.232 & 0.03304 & 3.0551  & 0.66531  & 28.016 & 0.9608 & 4.5920 \\
 $^{40}\mathrm{Ca}$  & 62.075 & 0.02558 & 4.2101  & 0.69815  & 53.067  &1.0251 & 6.0303 \\
 $^{48}\mathrm{Ca}$  & 61.339 & 0.03024 & 4.3166  & 0.66256  & 55.126  &1.2425 & 6.5150 \\
 $^{68}\mathrm{Ni}$    & 59.433 & 0.02748 & 4.9235  & 0.65209  & 69.486 &1.2876 & 7.5502 \\
 $^{90}\mathrm{Zr}$    & 59.981 & 0.02659 & 5.4609  & 0.67456  & 86.272 &1.3820 & 8.0955 \\
 $^{132}\mathrm{Sn}$ & 57.637 & 0.02662 & 6.13881 & 0.63767 & 104.76 &1.5553 & 9.6269 \\
 $^{208}\mathrm{Pb}$ & 57.996 & 0.02361 & 7.19934 & 0.68375 & 144.98 &1.61791 & 10.529 \\
 \hline
\end{tabular}
 \caption{Optimal central, spin orbit, half-density radius, and diffuseness Woods-Saxon parameters 
 	       for a representative set of doubly-magic nuclei. The parameters were fitted to effective 
	       Schr\"odinger-like neutron potentials derived from a realistic covariant energy density 
	       functional\,\cite{Chen:2014mza}. The last three columns list the corresponding dimensionless
	       parameters, as per Eqs.(\ref{Potential1}-\ref{Potential2}).}
 \label{Table1}
 \end{table*} 

Reduced basis methods will be compared against a conventional solution to Schr\"odinger's equation based on the 
Runge-Kutta algorithm. In this case, one uses the shooting method to obtain all bound states that are supported by 
the assumed mean-field potential. For the construction of the reduced basis, the following procedure is implemented. 
First, we start by generating ten random triplets for the three dimensionless parameters listed in Table\,\ref{Table1}, 
namely, $\lambda_{{}_{0}}$, $\lambda_{\rm so}$, and $\beta$. Given our goal of describing the entire single-particle 
spectrum for a variety of nuclei, those random values are drawn from a uniform distribution spanning the values listed 
in Table\,\ref{Table1}. For example, for the central potential, random values are drawn from a uniform distribution within 
the $28.02\le\lambda_{{}_{0}}\!\le144.98$ interval. Second, we train the reduced basis model by obtaining exact solutions 
for each of the ten realizations by invoking the Runge-Kutta method. In this manner, one generates a trained set of
(non-orthogonal) bound states for every angular momentum channel. Third, the optimal set of orthonormal basis functions 
is generated by filtering the (non-orthogonal) trained set through a singular value decomposition (SVD) routine. Finally, from 
such optimally generated set, one keeps the most important basis states as determined by their relative condition number.
That is, one only keeps those basis states for which the ratio of its singular value to the corresponding largest singular 
value exceeds the arbitrarily chosen bound of $10^{-3}$. This procedure generates reduced bases with dimensions 
ranging from as large as nine (for the $\kappa\!=\!-1$ sector) to as small as four (for the $\kappa\!=\!-7$ sector). Note 
that the binding energy is given in terms of dimensionless energy 
$\epsilon$ by $E_{\rm bind}\!=\!\epsilon V_{{}_{0}}/\lambda_{{}_{0}}$. It is important to underscore that by following such 
a procedure, we aspire to build a universal reduced basis that may be used without modification to generate all bound 
states---both ground and excited states---for all spherical nuclei. To our knowledge, the only extension of EC to excited 
states is the work by Franzke and collaborators that was applied to the one-dimensional quartic anharmonic 
oscillator\,\cite{Franzke:2021ofs}. 

\begin{figure}[ht]
 \centering
 \includegraphics[width=0.425\textwidth]{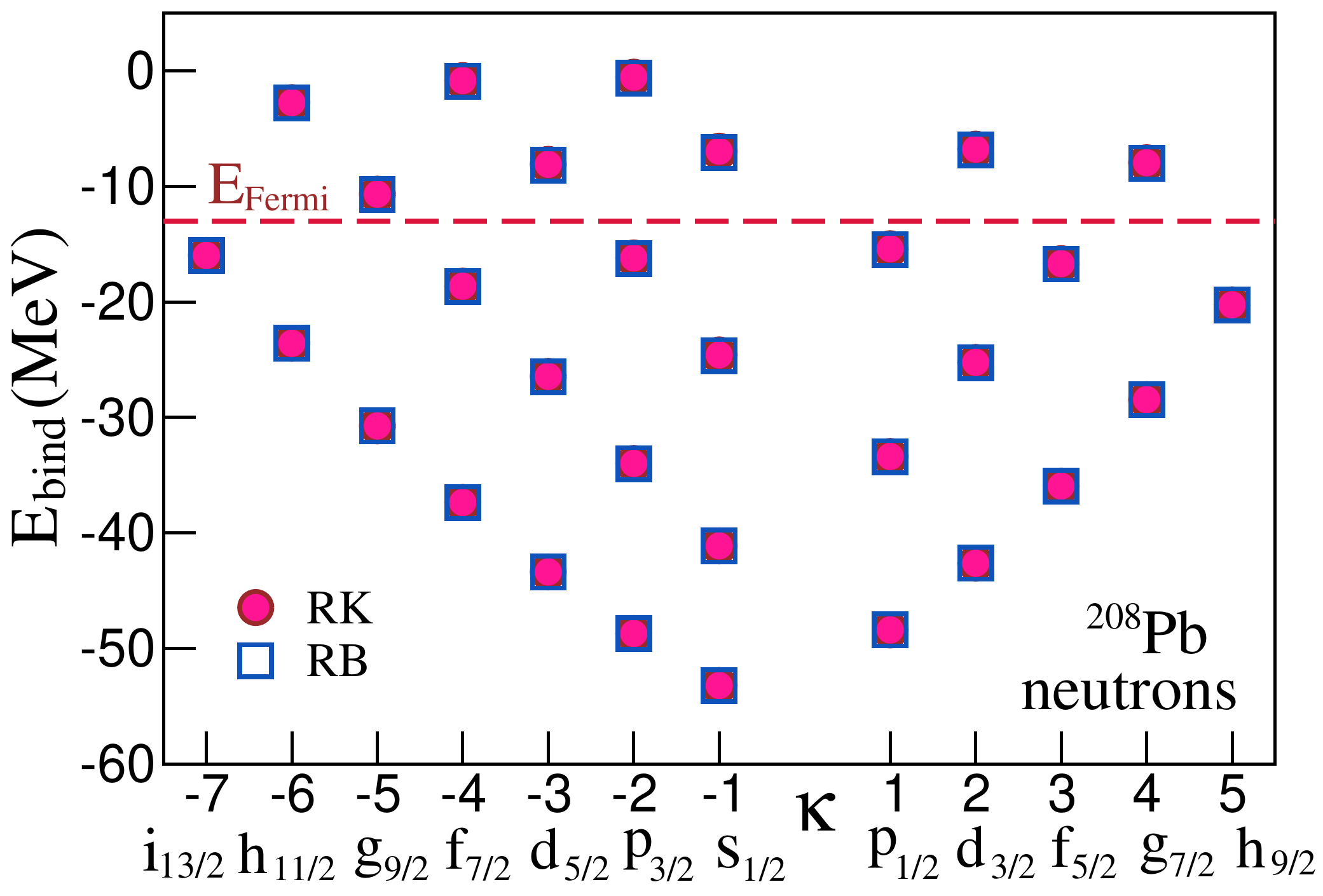}
 \caption{The entire single-neutron spectrum of ${}^{208}$Pb as generated by the mean-field potential listed
               in Table\,\ref{Table1}. The circles are the results obtained with the Runge-Kutta (RK) algorithm while  
               the squares are predictions using a well motivated reduced basis (RB). The dashed line indicates the 
               Fermi energy separating the 22 occupied orbitals from the eight empty ones.}
\label{Fig2}
\end{figure}

Besides the construction of a well-motivated basis that captures the essential physics, the reduced basis method also requires 
an efficient framework to solve the underlying set of dynamical equations. A powerful framework to do so, especially for systems 
of non-linear differential equations, is the Galerkin projection approach that determines the optimal set of expansion 
coefficients\,\cite{Quarteroni:2015,Bonilla:2022rph,Melendez:2022kid}. In our particular case, the Galerkin approach is equivalent
to the direct diagonalization of the Hamiltonian matrix in the reduced basis space. Note that the diagonalization of the Hamiltonian 
must be done within each angular momentum subspace.

We display in Fig.\,\ref{Fig2} the bound state neutron spectrum for ${}^{208}$Pb, generated by the exact Runge-Kutta 
algorithm (red circles) and the reduced basis method (blue squares). The agreement for both ground and excited states 
for all quantum numbers is excellent. Indeed, the root-mean-square error defined in terms of the total number of bound
states $N$,
\begin{equation}
 {\mathlarger\epsilon}_{\,\rm rms} = 
 \displaystyle\sqrt{\frac{1}{N}\sum_{n=1}^{N}\Big(\epsilon_{n}^{({\rm RK})}-\epsilon_{n}^{({\rm RB})}\Big)^{2}},
 \label{RMS}
\end{equation}
amounts to only ${\mathlarger\epsilon}_{\rm rms}\!\approx\!0.01$\,{\rm MeV}. The agreement is so good that one can
only provide approximate rms-errors, as it is unlikely that our codes can compute bound-state energies with a precision 
of better than 10\,keV. This suggests that the reduced basis could be made even smaller. 
The dashed line in Fig.\,\ref{Fig2} denotes the Fermi energy which divides the 22 occupied states 
from the 8 vacant (but still bound) single-particle orbitals. It is important to underscore that all states with the same 
angular momentum number $\kappa$---independent of the number of nodes---emerge directly from the diagonalization 
procedure within such an angular momentum sector. For example, in the $\kappa\!=\!-1$ ($l\!=\!0$, $j\!=\!1/2$) sector, 
the diagonalization of the $9\times 9$ Hamiltonian matrix yields exactly four bounds-state energies that are in very
close agreement with the predictions of the Runge-Kutta algorithm. 

\onecolumngrid
 \begin{center}
 \begin{figure}[ht]
  \includegraphics[width=0.99\textwidth]{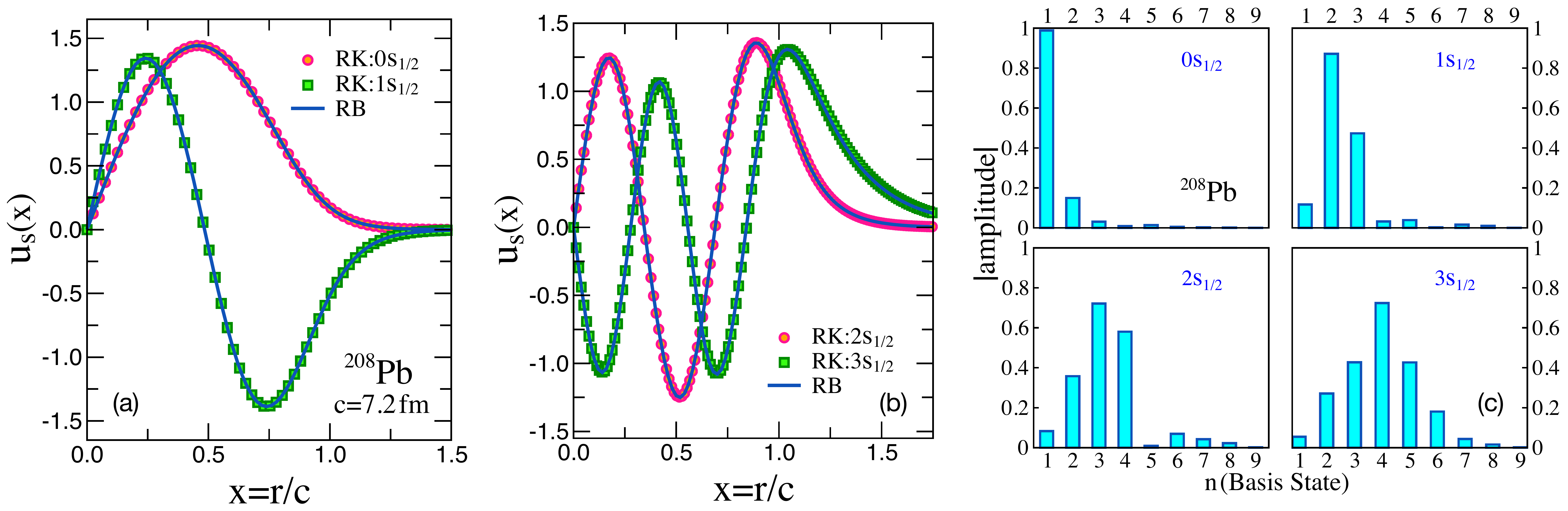}
  \caption{The first two (a) and the next two (b) {\large{s}}$_{1\!/\!2}$ bound-state orbitals of ${}^{208}$Pb as generated by 
  	        the Runge-Kutta algorithm (circles/squares) and the reduced basis method (solid lines). Also shown in (c)
	        are the absolute values of the projection amplitudes of all four states onto the $\kappa\!=\!-\!1$ reduced basis.}
 \label{Fig3}
\end{figure}
\end{center}
\twocolumngrid

Although the agreement is excellent, predictions for energies tend to be more accurate than for other observables by 
virtue of the Raleigh-Ritz variational principle. To ascertain that the success of the RBM goes beyond the bound-state 
energies, we have plotted in Fig.\,\ref{Fig3} the four $\kappa\!=\!-1$ bound-state wave-functions supported by the potential.
In the figure the Runge-Kutta results are depicted with circles/squares and the corresponding predictions from the RBM 
with solid lines. Not only are the wave-functions accurately reproduced, but the efficiency of the reduced basis is 
exceptional. Indeed, also displayed in Fig.\,\ref{Fig3} are the absolute values of the projection amplitudes of all four states 
onto the $\kappa\!=\!-1$ reduced basis. The figure illustrates how a well motivated reduced basis with only nine (or even 
seven!) states can accurately and efficiently reproduce the entire spectrum. In particular, the (nodeless) ground-state 
orbital can be essentially reproduced with only two basis states. To illustrate the universality of the basis, we now
demonstrate that the same reduced basis without any additional modification can reproduce the single-particle spectrum 
of ${}^{48}$Ca (with only 28 neutrons) as faithfully as in the case of ${}^{208}$Pb (with 126 neutrons). 

\begin{figure}[ht]
 \centering
 \includegraphics[width=0.425\textwidth]{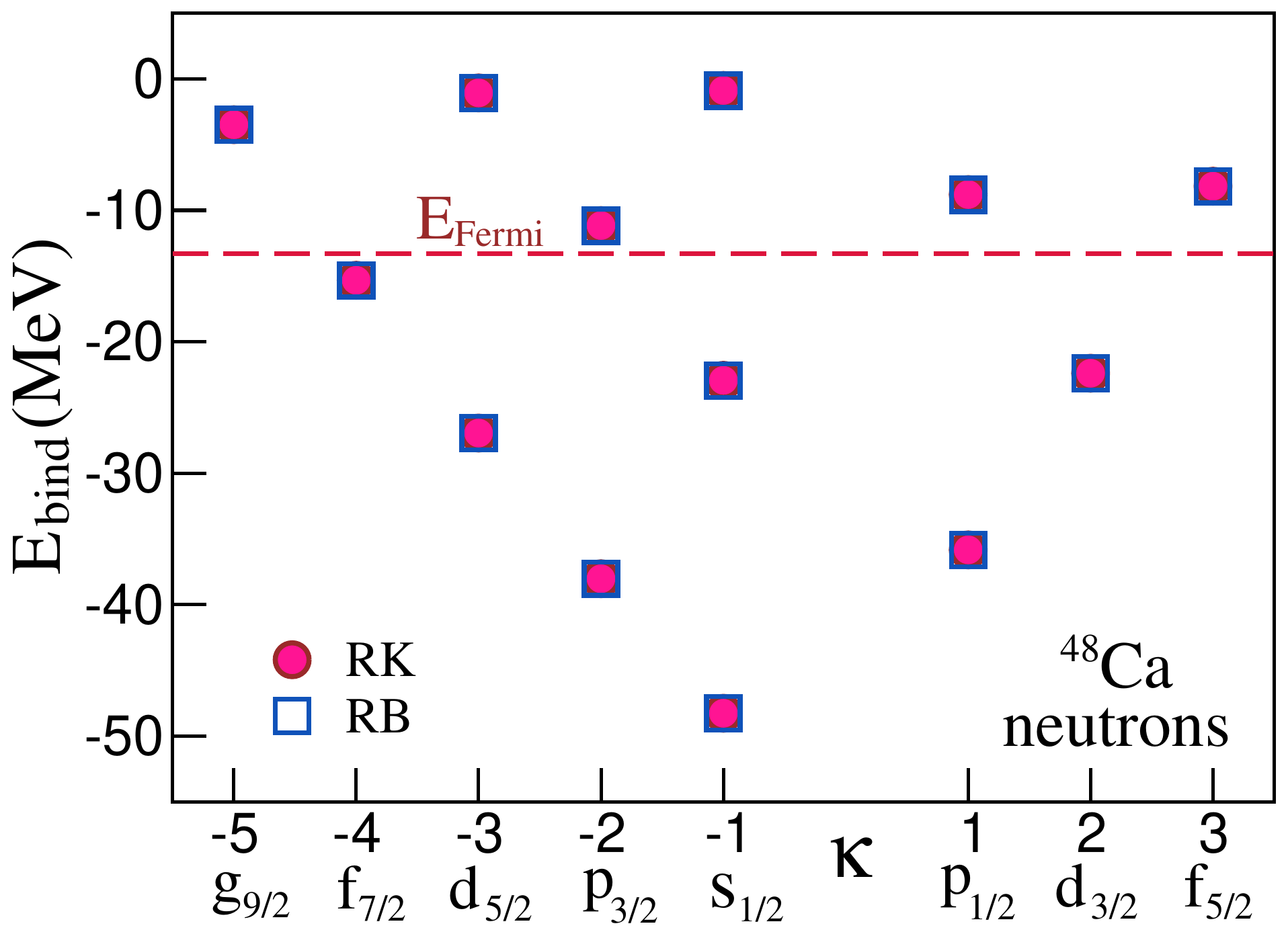}
 \caption{The entire single-neutron spectrum of ${}^{48}$Ca as generated by the mean-field potential listed
               in Table\,\ref{Table1}. The circles are the results obtained with the Runge-Kutta algorithm while the 
               squares are RBM predictions using a well motivated basis. The dashed line indicates the Fermi 
               energy separating the seven occupied orbitals from the six empty ones.}
\label{Fig4}
\end{figure}

\begin{figure}[ht]
 \centering
 \includegraphics[width=0.4\textwidth]{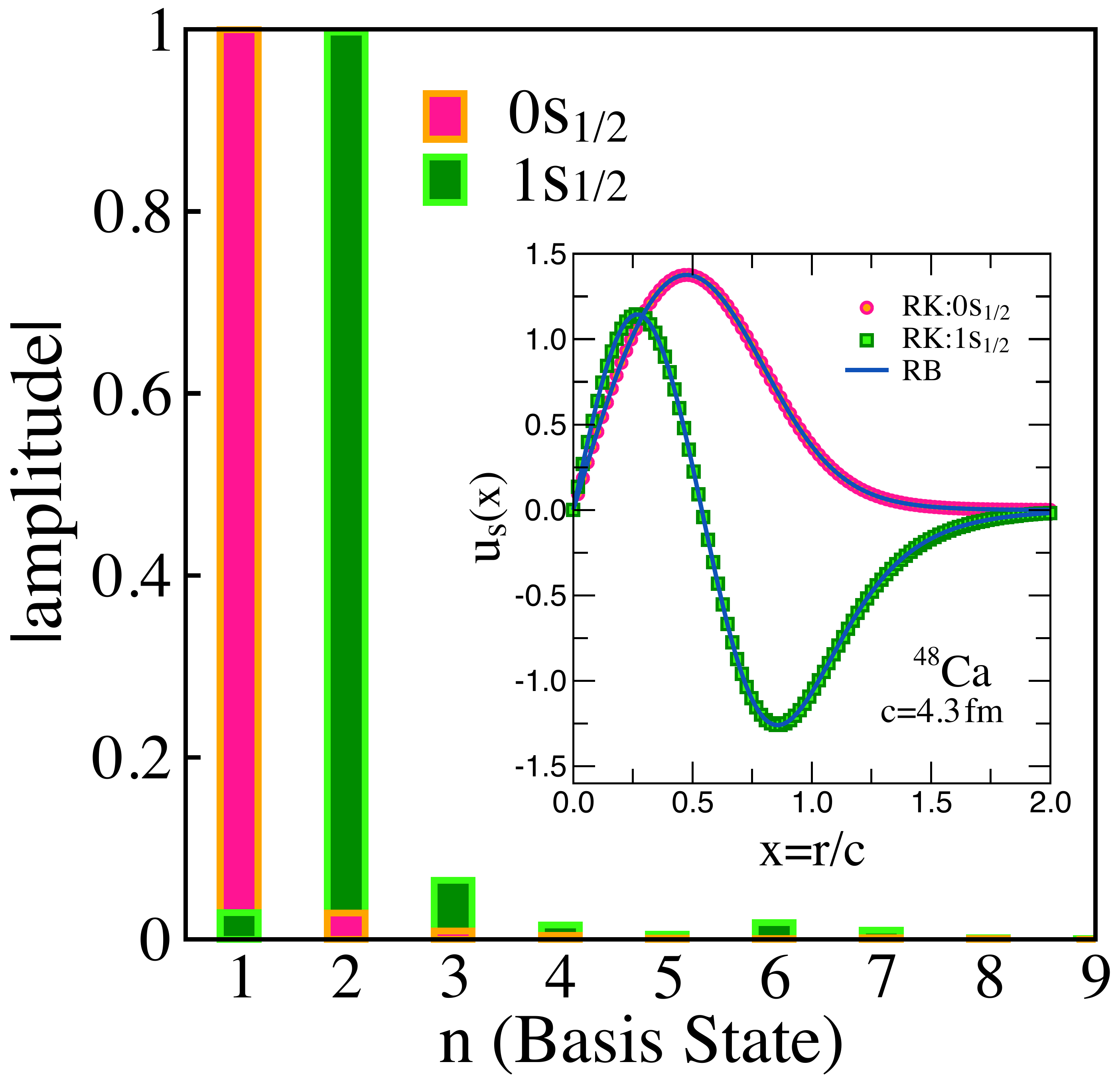}
  \caption{The two occupied {\large{s}}$_{1\!/\!2}$ bound-state orbitals of ${}^{48}$Ca as generated by 
  	        the Runge-Kutta algorithm (circles/squares) and the reduced basis method (solid lines). Also 
	        shown are the absolute value of the projection amplitudes of the two states onto the 
	        $\kappa\!=\!-\!1$ reduced basis.}
\label{Fig5}
\end{figure}

The entire single-neutron spectrum for ${}^{48}$Ca is shown in Fig.\,\ref{Fig4} using the same convention
as in the case of ${}^{208}$Pb. Again, the agreement between the exact results and the RBM predictions
is excellent, even for the barely bound states. In this case the rms error is about 
${\mathlarger\epsilon}_{\rm rms}\!\approx\!0.02$\,{\rm MeV}, although we reiterate that this is only an 
estimate since it is unlikely that we can compute bound-state energies with a precision of $\sim\!\!10$\,keV. 
We observe that the efficiency of the reduced basis in this case is even more impressive than for ${}^{208}$Pb. 
As shown in Fig.\,\ref{Fig5}, not only are the two occupied {\large{s}}$_{1\!/\!2}$ states in ${}^{48}$Ca accurately 
reproduced, but one can do so with essentially only one basis state, namely, with the one containing the 
correct number of nodes. 

In summary, we have constructed a highly efficient and accurate reduced order model to emulate the 
single-particle structure of atomic nuclei. The underlying mean-field Hamiltonian is constrained by the
predictions of a realistic energy density functional. 
The reduced basis was generated by filtering the trained set of eigenfunctions through a singular 
value decomposition routine. Although the universality of the reduced basis was demonstrated only 
for ${}^{48}$Ca and ${}^{208}$Pb, the accuracy of single-particle structure of all nuclei listed in 
Table\,\ref{Table1} was verified without any modification to the basis. These results will be 
presented in a forthcoming publication\,\cite{Anderson:2022}. To some extent, the success of the 
reduced basis method implemented here may be attributed to the simple scaling of the Hamiltonian. 
Although the half-density radius of the nuclei under consideration varies by more than a factor of two, 
the scaling leads to bound-state wave-functions that are highly similar to each other (see Figs.\,\ref{Fig3} 
and\,\ref{Fig5}) practically guaranteeing the robustness of the reduced basis and the success of the approach.   

In a longer publication we will explore further the universality of the reduced basis by computing both 
neutron and proton single-particle spectra for a variety of nuclei\,\cite{Anderson:2022}. We will quantify 
the speed performance of the RB emulator in comparison with the traditional RK solver. To achieve
significant gains, all the operators defining the dimensionless Hamiltonian should be linear in all the 
model parameters. As shown in Eqs.(\ref{Potential1}-\ref{Potential2}), the potential energy depends 
linearly on three of the four model parameters; the non-linear dependence is encoded in $\beta$, a 
relatively large quantity that is defined as the ratio of the half density radius to the diffuseness parameter. 
Hence, for the range of values of interest, the dimensionless Woods-Saxon form $f_{{}_{0}}(x;\beta)$
may be linearized using the Empirical Interpolation Method (EIM)\,\cite{Heasthaven:2016} as follows:
\begin{equation}
 f_{{}_{0}}(x;\beta)\!\approx\!\sum_{m=1}^{M} b_{m}(\beta) f_{m}(x),
 \label{EIM}
\end{equation}
where the functions $f_{m}(x)$ may be obtained following a procedure analogous to the one used to 
extract the reduced basis. That is, one generates Wood-Saxon potentials $f_{{}_{0}}(x;\beta)$ for different 
values of the parameters $\beta$ and then performs a singular value decomposition to identify and
retain the $M$ most important components. The nucleus-specific information is encoded in the 
coefficients $b_{m}(\beta)$ that may be determined by solving a set of $M$ linear equations after
identifying the various optimal locations $x$, usually chosen by a greedy algorithm 
scheme\,\cite{Heasthaven:2016}. Once the Hamiltonian has been fully linearized, matrix elements of 
the various components of the potential energy can be evaluated once (the offline stage) and then stored 
for later use (the online stage) to compute the spectrum of all nuclei of interest. Our ultimate goal is to build 
a reduced order framework to accurately and efficiently calibrate modern energy density functionals for which 
Bayesian optimization is demanded. This requires to compute the same observables for a large number of nuclei
many times to sample the entire parameter space in order to properly quantify correlations and model uncertainties. 
The computational burden is high, but it can be mitigated by using emulators which, as we have shown here, 
accurately and efficiently approximate the behavior of the original model, but at a highly reduced computational cost.

\begin{acknowledgments}\vspace{-10pt}
 We are grateful to Pablo Giuliani for his guidance, many stimulating discussions, and a thorough reading
 of the manuscript. This material is based upon work supported by the U.S. Department of Energy Office 
 of Science, Office of Nuclear Physics under Award Number DE-FG02-92ER40750. 
\end{acknowledgments} 

\vfill\eject

\bibliography{main.bbl}
\end{document}